\documentclass[draft]{agujournal2019}
\usepackage{url} 
\usepackage{lineno}
\usepackage[inline]{trackchanges} 
\usepackage{soul}
\usepackage{amsmath} 
\usepackage{amssymb}
\draftfalse

\begin{document}

\title{Overturning instability in forced ageostrophic oceanic flows}

\authors{Laur Ferris\affil{1}, Donglai Gong\affil{1}}
\affiliation{1}{Virginia Institute of Marine Science - William \& Mary, Gloucester Point, VA, United States}
\correspondingauthor{Laur Ferris}{lnferris@vims.edu}

\begin{keypoints}

\item The classical potential vorticity criterion ($qf<0$) neglects forced ageostrophic vertical shear, leading to missed overturning instability in forced ocean regimes.

\item We derive a coordinate-invariant criterion that incorporates along-front ageostrophic shear and recovers the PV criterion in the unforced limit, with direct application to model and observational datasets.

\item In an idealized ACC jet, the criterion alters the diagnosed depth structure and amount of overturning instability by up to 20\%.

\end{keypoints}

\begin{abstract}

The subpolar oceans are characterized by intense storm forcing and complex littoral topography. Submesoscale frontal instabilities are significant sources of turbulent kinetic energy (TKE) in these regions. However, criteria for identifying and parameterizing these instabilities in regional models have predominantly relied on a geostrophic framework that neglects generalized ageostrophic shear. We derive criteria for overturning instability that account for stabilizing and destabilizing effects of ageostrophic shear on mechanically forced boundaries, deviating from the geostrophically derived potential vorticity (PV) criterion, $qf < 0$. Ageostrophic forcing modifies stability from that implied by the vertical PV structure underlying bulk surface boundary layer diagnostics, which may limit the applicability of such bulk criteria in strongly forced regimes and motivate the need for layer-resolved measures. We demonstrate their application using a feature model of a wind-forced jet, as well as a 1-km Regional Ocean Modeling System (ROMS) hindcast of the high North Atlantic, and assess the importance of forced ageostrophic overturning instability (AOI) in intense frontal zones. In the feature model, ageostrophic shear increases overturning instability by up to 20\%, compared to a strictly geostrophic framework.
\end{abstract}

\section*{Plain Language Summary}
Subpolar oceans are home to strong storms, sharp density gradients, and complex coastlines that generate energetic ocean currents. Traditionally, oceanographers identify instability in these currents using an equation that ignores external forcing from wind and interactions with the seafloor. These forces are often present, causing the standard equation to miss unstable currents. To better detect instability, we developed an equation that explicitly accounts for the currents generated by these forces. We express the equation as a simple diagnostic that can be applied directly to model output and ocean observations. Our simulations suggest the new method identifies instability in different locations and depths, increasing detected instability by up to 20\% compared to the traditional approach. This improved representation of instability may help oceanographers improve predictions of upper-ocean structure and the performance of operational ocean models.

\section{Introduction}
Downscale energy transfer in the global ocean is an area of active research. While many processes are physically realizable, the predominant pathways of ocean energy transfer from the mesoscale to dissipative scale are still poorly understood. Above the thermocline, the energy cascade is powered by a combination of factors including wind stress, convective forcing, mesoscale eddies, submesoscale instabilities, and internal wave interactions. The multiple factors influencing the energy cascade produce complex energetic pathways leading to dissipation, most of which are difficult to quantify observationally. Understanding the physics of how energy moves downscale and produces mixing in the real ocean is necessary for the construction of next-generation operational and climate models, which currently utilize diffusivity-profile parameterizations based on vertical shear in the interior and Monin-Obukhov scaling near boundaries (e.g., KPP; \citeA{Large94}) or other prognostic TKE tendency equations (e.g. GLS/$k-\epsilon$/Mellor-Yamada; \citeA{Warner05}) to quantify turbulent mixing. With limited computational power, a critical task is elucidating the relative roles of the physical processes capable of generating and dissipating TKE.

The submesoscale includes flow features with spatial scales of $\mathcal{O}(1-10)$ km, vertical scales of $\mathcal{O}(100)$ m, timescales of $\mathcal{O}$(hrs-days), Richardson numbers $Ri =\mathcal{O}(1)$, and Rossby numbers $Ro = \mathcal{O}(1)$; and the effects of velocity structure, stratification, and rotation are all important. In a rotating, stratified ocean, overturning instabilities are found to arise when Ertel potential vorticity $q$, less its vertical velocity gradients, has a sign opposite of the Coriolis parameter $f$, i.e.~$qf<0$ \cite{Hoskins74}. These instabilities include baroclinic-symmetric (buoyancy gradient-induced), gravitational (stratification), and inertial/centrifugal (relative vorticity-induced) instabilities \cite{todd2016}. For gravitational instability (GI), energy is extracted from stratification ($B_z$) when $N^2 \equiv B_z <0$. For stratified shear instability (SSI), energy is extracted from vertical shear ($U_z,V_z$). For centrifugal instability (CI), energy is extracted from lateral shear ($U_y,V_x$) when $f\zeta_a <0$, where $\zeta_a = f +V_x-U_y$ is the absolute vorticity (see Table \ref{tab:table1} for a list of variables). Symmetric instability (SI) is a generalization of gravitational, shear, and centrifugal instability and can occur despite gravitational stability in the vertical and dynamic stability in the horizontal; energy is extracted by advecting vertical and lateral shear ($U_y,U_z,V_x,V_z$) of the along-front current \cite{Smyth19}. It can take on centrifugal-symmetric (CSI) or gravitational-symmetric (GSI) hybrid types; and arises from the same physical setup as baroclinic instability (along with a convective and an inertial critical layer mode) but assumes no along-front variation in the perturbation. 

\begin{figure}[htbp]
    \centering
    \includegraphics[width=0.6\textwidth]{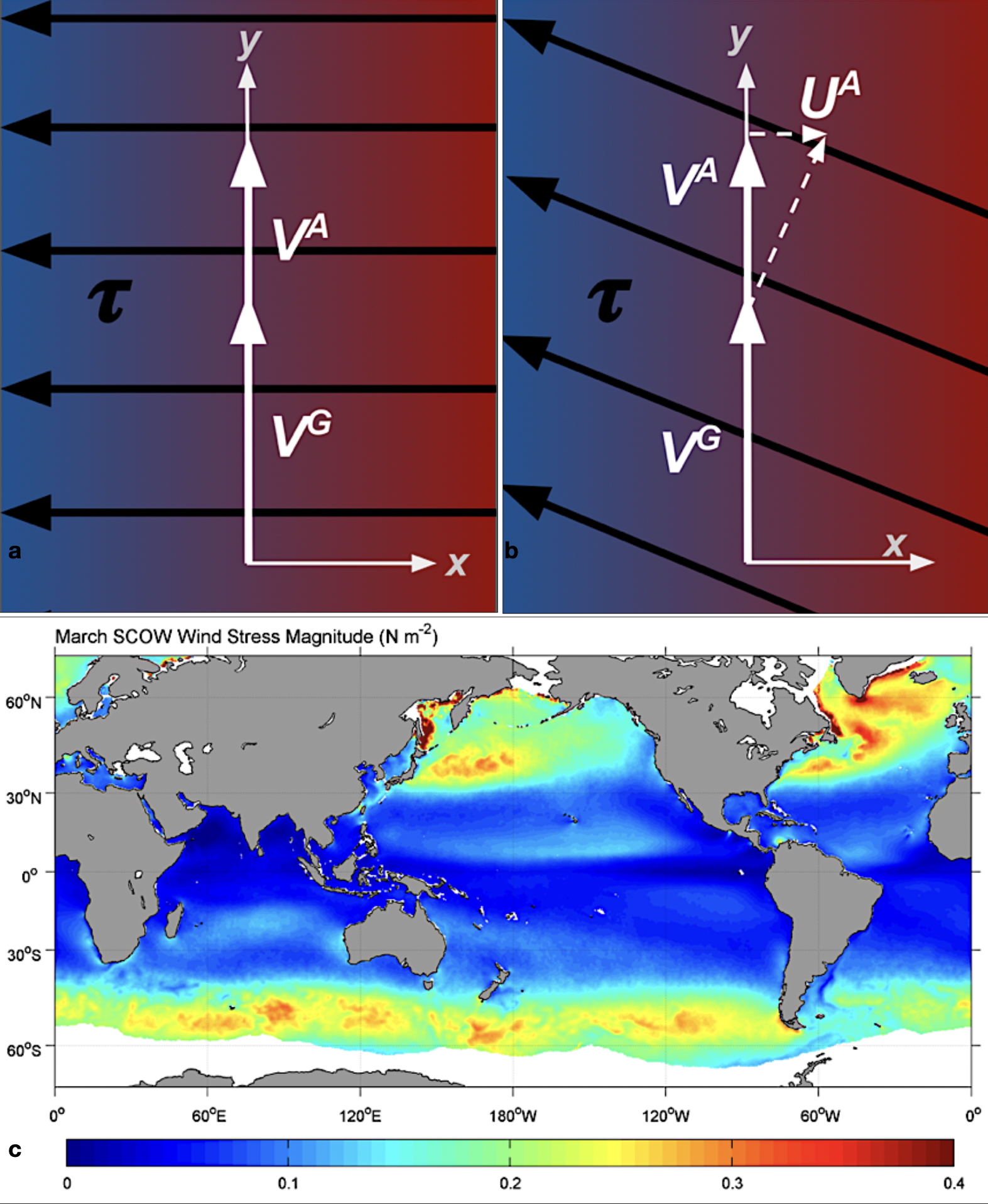}
    \caption{\textbf{(a)} Top-down diagram of an idealized baroclinic frontal region where the flow is a balanced between the Coriolis acceleration, horizontal pressure gradient force, and stress. Red indicates greater buoyancy. \textbf{(b)} Identical to (a) but illustrating a nonzero across-front ageostrophic velocity. \textbf{(c)} March wind stress from Scatterometer Climatology of Ocean Winds, map produced by \cite{Risien08}.}
    \label{fig:1}
\end{figure}

SI does not produce turbulence directly, rather, it produces weak restratification \cite{Dong21a} by extracting kinetic energy through shear production (see Fig. 3 of \citeA{Stamper17} for the velocity structure of growing SI) and thus can serve as conduit between the mesoscale, submesoscale, and the scale of TKE dissipation. Despite its potential importance for energy cascade from the ageostrophic submesoscale, the majority of prior investigations of SI have focused on its onset in geostrophic flows; studying the Eady problem in analytical and numerical situations \cite{Stone66,Haine98,Stamper17,Skyllingstad17,Skyllingstad20} and assuming thermal wind balance \cite{Boccaletti07,Thomas13,Thomas16,Adams17,NaveiraGarabato19} in observational settings.

Shear production by submesoscale instabilities in frontal zones has been found to have a significant contribution to global upper ocean turbulence, with typical mean values of $O(10^{-7})$ W kg$^{-1}$ (Dong et al 2024). Estimates for shear production by submesoscale instabilities (including SI) in frontal zones are also provided in (\citeA{Skyllingstad17} Fig. 6; \citeA{Thomas13} Fig. 9), which range $10^{-7}$ to $10^{-6}$ W/kg. Despite these non-trivial values, submesoscale instabilities are unresolved in most regional ocean circulation models because the wavelength of growing modes is typically 20-500 m \cite{Dong21a} and is unresolved in most regional ocean circulation models. Turbulence closure schemes (e.g., KPP, Generic Length Scale [GLS], Mellor-Yamada) generally consider vertical shear but neglect processes such as SI that are contingent on horizontal buoyancy and velocity gradients, for both boundary and interior treatments. 

An existing diffusivity parameterization that does incorporate SI \cite{Bachman17,Dong21a} focuses on geostrophic production by forced SI in the surface boundary layer (SBL). Forced SI occurs when wind (via Ekman buoyancy forcing) and meteorological buoyancy forcing sustain pre-existing SI. The parameterization relies on forced SI by checking for bulk negative potential vorticity in the SBL, attributing it to SI, and estimating geostrophic shear production. \citeA{Yankovsky21} have developed a parameterization for SI below the SBL which does not rely on dimensional parameters as does that of \citeA{Bachman17}, but it similarly leverages the assumption of thermal wind balance. Ageostrophic shear production \cite{Thomas13,Bachman17} is neglected for reasons both of simplicity and that its inclusion makes a negligible contribution to the energy budget when vertically integrated over the upper ocean \cite{Taylor10}. However, its depth-specific value can be a dominant term within the upper ocean (see Fig. 9 from \citeA{Thomas13}) and have a complex vertical structure compared to geostrophic shear production. Geostrophic methodology may be adequate for global or climatological models of geostrophic flows, but fall short in applications where accurate predictions of upper ocean structure are required. In summary, neither the effects of ageostrophy on the occurrence of instabilities, nor the ageostrophic shear production they may generate by SI are accounted for in current parameterizations of SI-associated turbulence and mixing. 

In this paper we generalize existing theory \cite{Stone66,Stone70,Hoskins74} to demonstrate that ageostrophic shear modifies the stability of upper ocean fronts, and thus can increase the shear production from submesoscale overturning instabilities. \citeA{Stone66,Stone70} established the existence and structure of non-geostrophic baroclinic instability in idealized settings, and \citeA{Hoskins74} provided a potential vorticity framework linking instability to PV gradients. We present a method for identifying overturning instabilities adapted from \citeA{Hoskins74} (which predates modern current profiling systems, when the assumption of geostrophic balance was essential) to represent the along-front effects of forced ageostrophy. We show this methodology alters the perceived stability of the flow using a feature model \cite{Gangopadhyay02} of the Antarctic Circumpolar Current (ACC); as well as a 1-km ROMS hindcast of the high North Atlantic \cite{Ferris25} developed during the Office of Naval Research (ONR)  Near Inertial Shear and Kinetic Energy in the High North Atlantic experiment (NISKINe).

\begin{table}[htbp]
  \centering
  \caption{List of variables}
  \label{tab:table1}
    \begin{tabular}{|l|l|}
        \hline
    $q^G$      & Ertel potential vorticity using only geostrophic velocity     \\     \hline
    $f$      & Coriolis parameter      \\    \hline
    $\rho_\theta$     & Potential density      \\    \hline
    $\rho_0$      & Reference density (1027 kg/m$^3$)      \\    \hline
    $x,y,z$      & Coordinates, z positive out of the ocean surface     \\    \hline
    $u,v,w,p$      & Perturbation velocity components and pressure      \\    \hline
    $U,V,P$     & Base state velocity components and pressure      \\    \hline
    $B$     & Buoyancy     \\    \hline
    $g$      & Acceleration due to gravity      \\    \hline
    $\psi$     & Velocity stream function     \\    \hline
    $\textbf{u}$     & Bold indicates a vector quantity     \\    \hline
    $N^2$     & Buoyancy frequency or stratification       \\    \hline
    $M^2$     & Lateral buoyancy gradient     \\    \hline
    $U_z$     & Subscript indicating a differentiated quantity
     \\    \hline
    $U^A_z$     & Superscript indicating only the ageostrophic part     \\    \hline
    $U^G_z$     & Superscript indicating only the geostrophic part     \\    \hline
    $\textbf{U}_{z\|}$    & \begin{tabular}{l}Subscript indicating a vector projection in the\\ direction of the front (positive downstream) \end{tabular}    \\    \hline
    $\zeta_a$    & Absolute vorticity     \\    \hline
    $\zeta_g$    & Absolute geostrophic vorticity     \\    \hline
    $Ri$     & Richardson number    \\    \hline
    $Ro$     & Rossby number    \\    \hline
    $\epsilon$     & Turbulent kinetic energy dissipation rate \\
        \hline
  \end{tabular}
\end{table}

\section{Theory}

 \subsection{Overturning instability criteria for an ageostrophic front}
 The most general criterion for overturning instability, $qf<0$, was derived in \citeA{Hoskins74}, after \citeA{Ooyama66} solution for the instability of symmetric disturbances in a baroclinic circular vortex. Ertel potential vorticity is given as follows:
\begin{equation}
q=(f \hat{k}+ \nabla \times \bold{u}) \cdot \nabla B = B_z(f+V_x- U_y) +B_yU_z -B_xV_z
\label{eq:1}
\end{equation}
Here, $B = -g\rho_{\theta}/\rho_0$. Both papers assume thermal wind balance, resulting in a criterion that applies for flows dominated by geostrophic dynamics. With the goal of studying instability in non-geostrophic flows, our first task is to derive Hoskin's criterion without the assumption of geostrophy. We begin with inviscid horizontal momentum equations for a perturbation in a jet (Fig. \ref{fig:1}a) in a rotating, stratified, forced system such that $\bold{U}= [0,V(x,z),0]$, where capitalized variables represent the base state and lowercase variables represent perturbation quantities.
\begin{subequations}
\begin{equation}
\frac{\partial u}{\partial t} + u \frac{\partial u}{\partial x} +(V+v)\frac{\partial u}{\partial y}+w\frac{\partial u}{\partial z}  -f(V+v)= -\frac{1}{\rho_0}\frac{\partial}{\partial x}(P+p)+\Upsilon
\end{equation}
\begin{equation}
\frac{\partial v}{\partial t} + u \frac{\partial(V+v)}{\partial x} +w\frac{\partial (V+v)}{\partial z} +fu= 0
\end{equation}
\label{eq:2}
\end{subequations}
Hoskins (1974) concluded that generation of SI in a previously stable flow \textit{requires} external forcing such as friction or heating. Similar to \citeA{Hoskins74} we consider the case where there is no variation in the along-front ($y$) direction, and the base state does not curve or evolve in time. We do not require the flow be in geostrophic balance; the final term in (Eq. \ref{eq:2}a) represents mechanical forcing $\Upsilon(x,z)$ such as a pressure drag, or a wind stress acting over the surface boundary layer in the x-direction e.g., $ \Upsilon(x,z) = \rho_0^{-1}\partial \tau^{\hat{x}}/\partial z$, where $\tau(x,z)$ is the stress acting in the fluid. This permits an ageostrophic balance between the pressure gradient force, Coriolis parameter, and external forcing such that $V = V^G+V^A$ (Fig. \ref{fig:1}a). Noting $-fV^G = -\rho_0^{-1}\partial P/\partial x$ and $-fV^A = \Upsilon$, our system representing a forced ageostrophic jet reduces to the same linearized, Boussinesq, inviscid, adiabatic equations of \citeA{Ooyama66} and \citeA{Hoskins74}, which represented unforced geostrophic jets. 

This derivation incorporates the direct modification of the along-front shear ($V_z$) by ageostrophic flow; but we briefly comment on the significance of a nonzero across-front ageostrophic velocity (Fig. \ref{fig:1}b), which is identically absent from the $qf<0$ criterion. In a geostrophic framework SI is defined as a frontal instability arising from the interaction of rotation with the lateral and vertical shear of the along-front velocity ($V_x,V_z$) \cite{Smyth19}; where across-front ageostrophic flow ($U$) is considered separately for its influence in perpetuating or eliminating the symmetrically unstable setup \cite{Hamlington14,Skyllingstad17}. If phenomena associated with \textit{across-front} ageostrophic flow are considered to be \textit{part of} SI and the hybrid suite of front-associated instabilities, this derivation would need to include vertical perturbation of the across-front shear ($wU_z^A\hat{x}$) and advection of the perturbation and along-front velocity ($U^Au_x\hat{x},U^AV_x+U^Av_x\hat{z}$), with the lateral density gradient, by the across-front ageostrophic velocity; potentially at the expense of an analytical solution.

The scope of this paper also does not consider an unbalanced ageostrophic flow (nonzero material derivative) or the effects of curvature \cite{Shakespeare16,Buckingham21a,Buckingham21b}, which are important at low latitudes or increasingly small scales. However, the role of externally forced ageostrophy, alone, in the global ocean is non-negligible (Fig. \ref{fig:1}c). For example, a wind stress of $\tau=0.3$ N/m$^2$ over a 20-meter Ekman layer produces an Ekman velocity of O(0.15m/s) --- an ageostrophic velocity contribution of 10-30\% in a frontal zone with base state velocity of 0.5-1.5 m/s. AOI in unbalanced or curving ageostrophic flows must be a subject of follow-on research.

We linearize the system in Eq. \ref{eq:2}, taking the jet to be in quasi-equilibrium, and assume negligible along-front variation (over the scale of calculation) to obtain perturbation equations. In linearizing we assume the background flow evolves more slowly than the timescale of developing instability itself.
\begin{subequations}
\begin{equation}
\frac{\partial u}{\partial t} -fv = -\frac{1}{\rho_0}\frac{\partial p}{\partial x}
\end{equation}
\begin{equation}
\frac{\partial v}{\partial t}+u\frac{\partial V}{\partial x}+w\frac{\partial V}{\partial z}+fu = 0
\end{equation}
\begin{equation}
\frac{\partial w}{\partial t} = - \frac{1}{\rho_0}\frac{\partial p}{\partial z}-g\frac{\rho_\theta}{\rho_0}
\end{equation}
\begin{equation}
\frac{\partial \rho_\theta}{\partial t}+u\frac{\partial \rho_\theta}{\partial x}+w\frac{\partial \rho_\theta}{\partial z}=0
\end{equation}
\label{eq:3}
\end{subequations} 

We define streamfunction $u = -\partial \psi/\partial z$ and $w=\partial \psi/\partial x$, lateral bouyancy gradient $M^2 = (-g/\rho_0)(\partial \rho_{\theta}/\partial x)=B_x = fV_z^G $, stratification $N^2 = (-g/\rho_0)(\partial \rho_{\theta}/\partial z)=B_z$, and vertical shear $n^2 = fV_z$. 
We apply $-\partial/\partial z$  to (Eq. \ref{eq:3}a) and $\partial/\partial x$ to (Eq. \ref{eq:3}c). Differentiating their sum by $t$, and substituting (Eq. \ref{eq:3}d) and (Eq. \ref{eq:3}b) we obtain
\begin{subequations}
\begin{equation}
\begin{split}
\frac{\partial ^2}{\partial t^2}\Big(\frac{\partial^2}{\partial x^2} + \frac{\partial^2}{\partial z^2}\Big) \psi+\Big[N^2\frac{\partial^2}{\partial x^2}-(n^2+M^2)\frac{\partial^2}{\partial x \partial z}+f\zeta_a\frac{\partial^2}{\partial z^2}\Big] \psi+C = 0 \\
\text{ where }
C = \frac{\partial \psi}{\partial x}\frac{\partial N^2}{\partial x}-\frac{\partial M^2}{\partial x}\frac{\partial \psi}{\partial z}+f\frac{\partial \psi}{\partial z}\frac{\partial^2V}{\partial z \partial x}- f\frac{\partial \psi}{\partial x}\frac{\partial^2V}{\partial z^2}
\end{split}
\end{equation}
Noting $\partial M^2/\partial z = \partial N^2/\partial x$, we obtain $C = -f(\bold{u}\bullet\nabla)\partial V^A/\partial z$; thus, $C$ is the influence of inflections. Comparing to (Eq. 19 from Ooyama, 1966; reproduced in Hoskins, 1974), we note that an assumption of thermal wind balance would mean $n^2 \equiv M^2$, recovering the geostrophic form of this equation:
\begin{equation}
\frac{\partial^2}{\partial t^2}\Big(\frac{\partial^2}{\partial x^2} + \frac{\partial^2}{\partial z^2 } \Big )\psi + \Big (N^2\frac{\partial ^2}{\partial x^2} - 2M^2 \frac{\partial ^2 }{\partial x \partial z}+f\zeta_a \frac{\partial^2 }{\partial z^2}  \Big)\psi   =  0
\end{equation}
\label{eq:5}
\end{subequations}
A consequence of the departure of velocity shear from geostrophy, $n^2-M^2$, is a deviation of the stability criterion from the $qf<0$ criterion derived in \citeA{Hoskins74}. 

\begin{table}[htbp]
  \centering
  \caption{Derivatives in normal mode form}
  \label{tab:table2}
  \begin{tabular}{|l|l|l|}
    \hline
    $\frac{\partial }{\partial x} = ik\sin{\phi}$ & $\frac{\partial }{\partial z} = ik\cos{\phi}$ &    \\ \hline
    $\frac{\partial^2 }{\partial x^2} = -k^2\sin^2{\phi}$ & $\frac{\partial^2 }{\partial z^2} = -k^2\cos^2{\phi}$ & $\frac{\partial^2 }{\partial x\partial z}= -k^2\sin{\phi}\cos{\phi}$     \\ \hline
    $\frac{\partial^2 }{\partial t^2}\frac{\partial^2 }{\partial x^2} = \sigma^2k^2sin^2{\phi}$ & $\frac{\partial^2 }{\partial t^2}\frac{\partial^2 }{\partial z^2} = \sigma^2k^2\cos^2{\phi}$ &  \\ \hline
  \end{tabular}
\end{table}
The derivatives of normal mode solution $\hat{\psi} = e^{i\sigma t}e^{ik(x\sin\phi+z\cos\phi)}$, where $\phi$ is the horizontal angle of the disturbance and $i\sigma$ is the complex growth rate, are given in Table \ref{tab:table2}. Letting $\tau = \tan\phi$ after Hoskins (1974), Eq. \ref{eq:5}a becomes:     
\begin{equation}
\begin{split}
\sigma^2 = cos^2\phi [N^2\tau^2  -(M^2+n^2)\tau + f\zeta_a + i\hat{C}]  \\
\text{ where }  i\hat{C}=C(\hat{\psi}) = i(n^2-M^2)_x\cos^{-1}\phi + i\tau(M^2-n^2)_z\cos^{-1}\phi 
\end{split}
\end{equation}
The normal mode solution experiences exponential unstable growth when $\text{Im}\{\sigma\} < 0$, or $\sigma^2 <0$, which occurs for some $\phi$ when the discriminant of the bracketed term indicates two real quadratic roots.

The normal mode form requires coefficients do not depend on $(x,z)$ such that we neglect $i\hat{C}$, approximating a condition of constant shear \textit{within each discrete layer} to which the resulting criterion is applied. In other words, the inflection of ageostrophic shear ($V_{zz}^A$ and $V_{zx}^A$) must be neglected over the vertical scale of calculation, $\Delta z$, due to the $x$ and $z$ independence assumed by $\psi$; a concession made to facilitate analytical solution. With discriminant $\Delta = b^2-4ac$ and $a = N^2$, $b = -(M^2+n^2)$, and $c = f\zeta_a$, the condition for unstable modes to exist becomes
\begin{equation}
M^4+2M^2n^2+n^4-4N^2f\zeta_a>0
\label{eq:6}
\end{equation}
In the absence of ageostrophic shear, Eq. \ref{eq:6} reduces to the familiar potential vorticity expression for instability in a meridional geostrophic flow \cite{Hoskins74}:
\begin{equation}
0 > fB_z(f+V_x)-fB_xV_z^G = q^Gf
\label{eq:7}
\end{equation}
A criterion $qf<0$ applied using the general form of potential vorticity (Eq. \ref{eq:1}) is sometimes called the ``potential vorticity (PV)" or ``total PV" criterion, while $q$ comprised using only geostrophic vertical shears ($q \equiv q_G$ as in Eq. \ref{eq:7}) is often called the ``geostrophic" or``balanced" criterion. Neither is the most appropriate if ageostrophic shear is meaningfully present. The condition for instability in a flow experiencing balanced ageostrophic shear (e.g. from wind forcing or flow-topography interaction) is obtained by expressing Eq. \ref{eq:6} in terms of absolute velocity and buoyancy gradients:
\begin{equation}
0 > fB_z(f+V_x)-[B_x^2+2fB_xV_z+f^2V_z^2]/4 
\label{eq:8}
\end{equation}
We can express (Eq. \ref{eq:8}) as:
\begin{equation}
 0 > fB_z(f+V_x)-fB_xV_z-f^2V_z^{A2}/4 =fq - f^2 V_z^{A2}/4 
\label{eq:9}
\end{equation}
The difference between the ``total PV" interpretation of the $qf<0$ criterion and the ageostrophic criterion is the negative term in (Eq. \ref{eq:9}); however, ageostrophic shear is not necessarily ``destabilizing" (i.e. admits additional unstable modes) because it also influences $q$. (Rearranging Eq. \ref{eq:8}, we find ageostrophic shear $V_z^A$ is destabilizing when it has the same sign as the geostrophic shear $V_z^G$, or opposes it by more than four times the magnitude, e.g. $V_z^GV_z^A + V_z^{A2}/4>0$.) 

\subsection{Generalization to arbitrary front orientations}
\begin{figure}[htbp]
    \centering
    \includegraphics[width=0.3\textwidth]{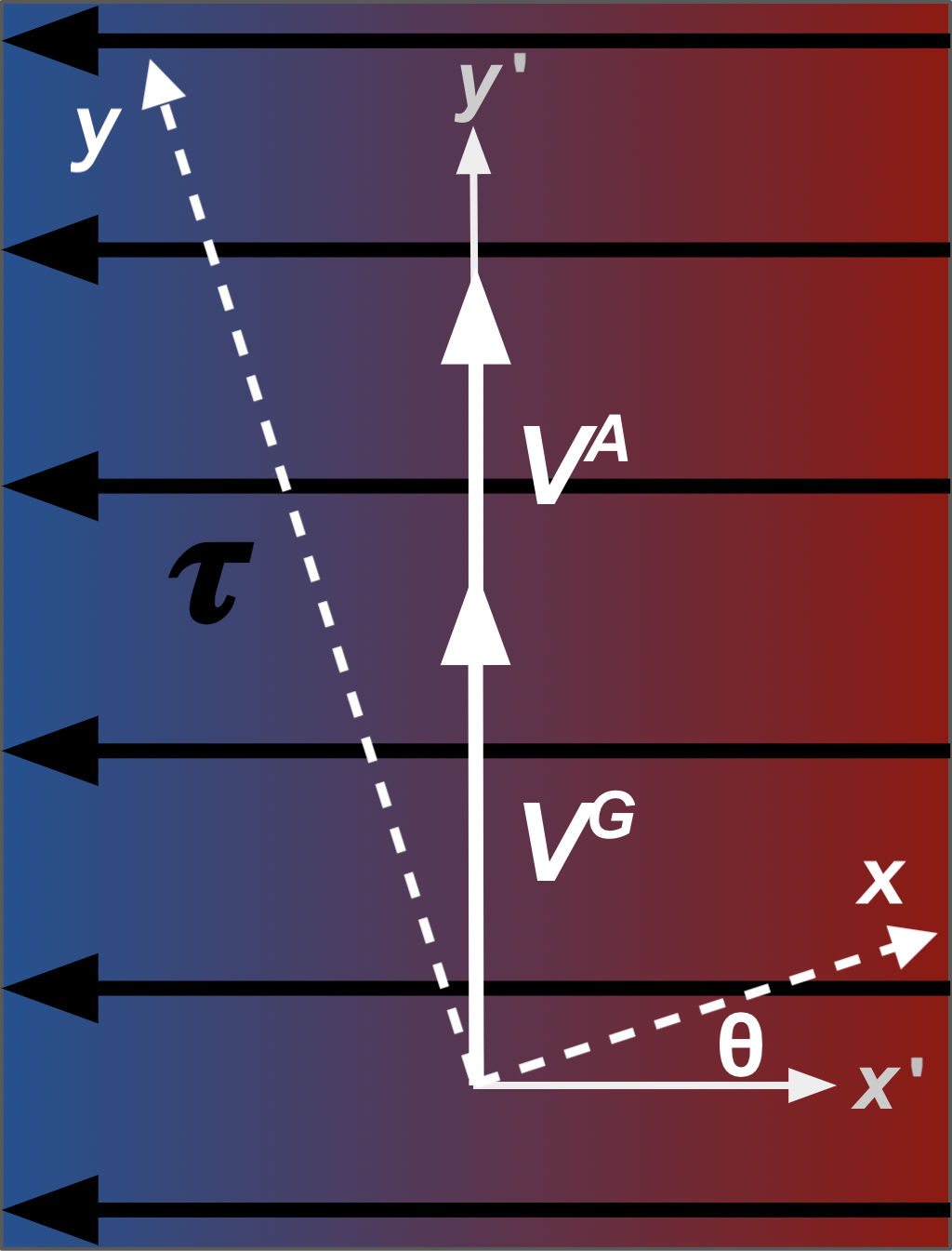}
    \caption{Coordinate redefinition of Fig. \ref{fig:1}ab to validate generalization of Eq. \ref{eq:8} to Eq. \ref{eq:10}.}
    \label{fig:2}
\end{figure}
The final task is to generalize the criterion (Eq. \ref{eq:8}) for instability in a front orthogonal to the Cartesian coordinate system (Fig. \ref{fig:1}ab) to a jet of any orientation. For a perturbation symmetric with respect to the local front direction, rotating into front-aligned coordinates reduces the stability problem to Eqs. 4a-5. The coordinate-invariant expression of the resulting instability condition is:
\begin{equation}
\begin{split}
0>fB_z(f+V_x-U_y)-fA \\	                
 \text{ where } A = [(B_x^2+B_y^2)/f-2B_yU_z+2B_xV_z+f(U_{z||}^2+V_{z||}^2)]/4
\end{split}
\label{eq:10}
\end{equation}
The new terms in $A$ encompass destabilization due to perturbation of the absolute vertical shear, $(U_{z\|}^2+V_{z\|}^2)$ and lateral buoyancy gradient, $(B_x^2+B_y^2)$. The general potential vorticity criterion (Eq. \ref{eq:1}) is recovered upon equating geostrophic shear with absolute velocity shear, $V_z \equiv V_z^G=B_x/f$ and $U_z \equiv U_z^G=-B_y/f$.

The final term contains horizontal projections of vertical shear in the direction of the front, $\bold{U}_{z\|}=\bold{U^G_z}(U_z U_z^G+V_zV_z^G)/(U_z^{G2}+V_z^{G2})$, which arise when generalizing (Eq. \ref{eq:8}) to a flow where ageostrophic shear is not aligned with geostrophic shear. Remaining velocity shear terms in (Eq. \ref{eq:10}) are unchanged because projection of absolute velocity shear in the direction of geostrophic velocity is the rejection of absolute velocity shear in the direction of lateral buoyancy gradient, $f(U_zU_z^G+V_zV_z^G)/\|\nabla B \| =(V_zB_x-U_zB_y)/\|\nabla B\|$; they are projected along the front by definition. This is verifiable by letting $\bold{a}=\langle U_z,V_z \rangle	$ and $\bold{b}=\langle fU_z^G,fV_z^G \rangle$, and noting $\bold{a}_\|\bullet\bold{b} = \bold{a}\bullet\bold{b}$.

The generalization of Eq. \ref{eq:8} into Eq. \ref{eq:10} is validated by redefining the coordinate system aligned with the front to be $x'-y'$, and a generalized coordinate system not aligned with the front, $x-y$ (Fig. \ref{fig:2}); with the goal of showing Eq. \ref{eq:10} defined on $x-y$ is equivalent to Eq. \ref{eq:8} defined on $x'-y'$. Vector components are related:
\begin{equation}
\bold{X} = 
\begin{bmatrix}
\cos{\theta} & \sin{\theta} \\
-\sin{\theta} & \cos{\theta} 
\end{bmatrix} \bold{X'}                                                                                                         
\label{eq:11}
\end{equation}
where $\bold{X}$ is $\langle B_x,B_y \rangle$ or $\langle U_z,V_z \rangle$. The first term in Eq. \ref{eq:10}, $fB_z(f+V_x-U_y)$, is rotationally invariant as property of absolute vorticity, and so this task simplifies to showing equivalence of the other terms in Eq. \ref{eq:10} and Eq. \ref{eq:8}. In (Fig. \ref{fig:2}) coordinates, $fA$ (Eq. \ref{eq:10}) becomes:
\begin{equation}
 [ B_x'^2+B_y'^2 +
2fB_x'V_z'-2fB_y'U_z'+f^2(U_z' U^G_z\prime +V_z' V_z^G\prime)^2/(U_z^G\prime^2 +V_z^G\prime^2)]/4
\label{eq:12}
\end{equation}
\begin{table}[htbp]
    \centering
    \small   
    \caption{Criteria for overturning instability}
    \label{tab:table3}
    \begin{tabular}{|l|l|l|}
    \hline
    \textbf{Criterion} & \textbf{Comments} & \textbf{Equation} \\
    \hline
    \begin{tabular}{l}  Geostrophic\\ overturning\\ instability\\ (GOI)\end{tabular} &\begin{tabular}{l}Eq. 7: Hoskins (1974)\\ Eq. 13a: Thomas et al. (2013)\end{tabular} 
   & \begin{tabular}{l}$0>qf$ where\\ $q= B_z(f+V_x-U_y)+B_yU_z^G-B_xV_z^G$\end{tabular}  \\
    \hline
    \begin{tabular}{l}Overturning \\instability\\ (OI) \end{tabular}
    & \begin{tabular}{l}Community interpretation of\\ Hoskins (1974) and Thomas\\ et al. (2013); e.g. Naveira\\ Garabato, et al. (2019) \end{tabular}
    &\begin{tabular}{l} $0>qf$ where\\ $q= B_z(f+V_x-U_y)+B_yU_z-B_xV_z$\end{tabular}  \\
    \hline
   \begin{tabular}{l}Ageostrophic\\ overturning\\ instability\\ (AOI)\end{tabular} & \begin{tabular}{l}Eq. 8: Carpenter et al. (2020)\\ Eq. 9: Haney et al. (2015)\\ Eq. 10: This paper \end{tabular} 
   &\begin{tabular}{l} $0>fB_z(f+V_x-U_y)-fA$ where\\ $A = [(B_x^2+B_y^2)/f-2B_yU_z+2B_xV_z+f(U_{z||}^2+V_{z||}^2)]/4$\end{tabular}  \\
    \hline
    \end{tabular}
    \end{table}
Substituting $B_y' = -fU_z^G\prime = 0$ as in the original setup confirms Eq. \ref{eq:10} is a generalization of Eq. \ref{eq:8}.

To summarize, Eq. \ref{eq:10} is the condition for overturning instability derived without the assumption that along-front shear results from leading-order geostrophic balance, and is expressed in terms of absolute velocity components. Forms of Eq. \ref{eq:8} and Eq. \ref{eq:9} have appeared previously (e.g., \citeA{Carpenter20}, Text S3; \citeA{Haney15}, Eq. 31). Here we show that Eq. \ref{eq:10} constitutes the condition for overturning instability in the presence of ageostrophic shear, and express it in a coordinate-invariant form that can be evaluated directly in arbitrary flows without requiring prior identification of frontal orientation, enabling practical application to gridded model output and observations.

\subsection{Practical criteria that accommodate along-front ageostrophic shear}

Hoskins (1974) recast the general $qf<0$ inequality in terms of the balanced Richardson number $Ri = B_z/(U_z^2+V_z^2)$ by using only geostrophic part of vertical shears in (Eq. \ref{eq:1}), such that instabilities arise when
\begin{subequations}
\begin{equation}
Ri_B = N^2/(U_z^{G2}+V_z^{G2}) \equiv f^2N^2/|\nabla_hb|^2 <f/\zeta_g
\end{equation}
Here $f\zeta_g <0$ and $\zeta_g$ is the absolute geostrophic vorticity and superscripts indicate geostrophic shear. Use of absolute geostrophic vorticity rather than absolute vorticity is not mathematically necessary (see Section 1 or \citeA{Haine98} or \citeA{Smyth19}). \citeA{Thomas13} supposes that the flow is sufficiently dominated by geostrophic dynamics such that $\zeta_g \approx \zeta_a$; we similarly suppose horizontal shear is dominated by geostrophic dynamics, and retain the symmetric property $\zeta_a \approx \zeta_g$. Overturning instabilities arise when $\Phi_{Ri_B}<\Phi_c$, where $\Phi_{Ri_B} = \tan^{-1}(-1/Ri_B)$ and $\Phi_c = \tan^{-1}(-\zeta_g/f)$. The angles can be used for the compact identification of instability types (see \citeA{Thomas13} for further detail). This expression is advantageous in two situations: (1) when one is interested in the basin-scale geostrophic behavior of a flow, and (2) when one lacks velocity data from a model or ADCP and must estimate velocity shear from $\nabla B$. The practical utility of \citeA{Thomas13} is retained, while admitting some ageostrophic effects, by rearranging Eq. \ref{eq:10}:
\begin{equation}
Sr = fN^2/A<f/\zeta_a
\end{equation}
\label{eq:13}
\end{subequations}
We use the dimensionless placeholder $S_r$ (in analogy to $Ri_B$) to relate the influences of stratification and shear. Excluding barotropic CI, overturning instabilities arise when $\Phi_{Sr}<\Phi_c$, where $\Phi_c = \tan^{-1}(-\zeta_a/f)$ and $\Phi_{Sr} = \tan^{-1}(-1/Sr)$; the latter of which must be implemented as a four-quadrant inverse tangent function e.g. $\text{atan2d}(-A/f,B_z)$. Their subtypes are given in Table \ref{tab:table4}, and all variables are summarized in Table \ref{tab:table1}. 
\begin{table}[htbp]
    \centering
    \caption{Criteria for hybridized subtypes of overturning instability}
    \label{tab:table4}
    \begin{tabular}{|l|l|}
        \hline
        \textbf{Type} & \textbf{Criteria} \\
        \hline
        Gravitational Instability (GI) & $\Phi_{Sr} < -135$ \\
        \hline
        Gravitational-Symmetric Instability (GSI) & $ -135 < \Phi_{Sr} < -90$ \\
        \hline
        Symmetric Instability (SI) &     \begin{tabular}[t]{@{}l@{}}
        $-90 < \Phi_{Sr} < \Phi_c$  and $\Phi_c < -45$ \\
        \textbf{or} \\
        $-90 < \Phi_{Sr} < -45$ and $-45 < \Phi_c$
        \end{tabular} \\
        \hline
        Centrifugal-Symmetric Instability (CSI) & $-45 < \Phi_{Sr} < \Phi_c$  and $-45 < \Phi_c$\\
        \hline
    \end{tabular}
\end{table}

Forcing from wind stress, buoyancy flux, and wave effects can produce a variety of ageostrophic overturning instabilities which may not be consistent with those classic for submesoscale fronts. When analyzing model output we suggest excluding $0 < Ri < 0.25$, where AOI with symmetric qualities may exist with little meaning, e.g., is dominated by Kelvin-Helmholtz instability or mixed shear-convective instability. We also note that all versions of the criteria (Table \ref{tab:table3}) were derived via quasi-equilibrium theory, or the ``frozen flow" approximation, such that their validity holds only when the background flow ($B,U,V$) does not adjust on a timescale comparable to the growth rate ($\sigma$) of the instability. Thus, they may not be appropriate for direct application to some regions and models; e.g. large eddy simulations (LES) of rapidly evolving flows. 

\section{Example Cases}
\subsection{Application to an idealized analytical model of a jet}
 \begin{figure}[htbp]
    \centering
    \includegraphics[width=0.9\textwidth]{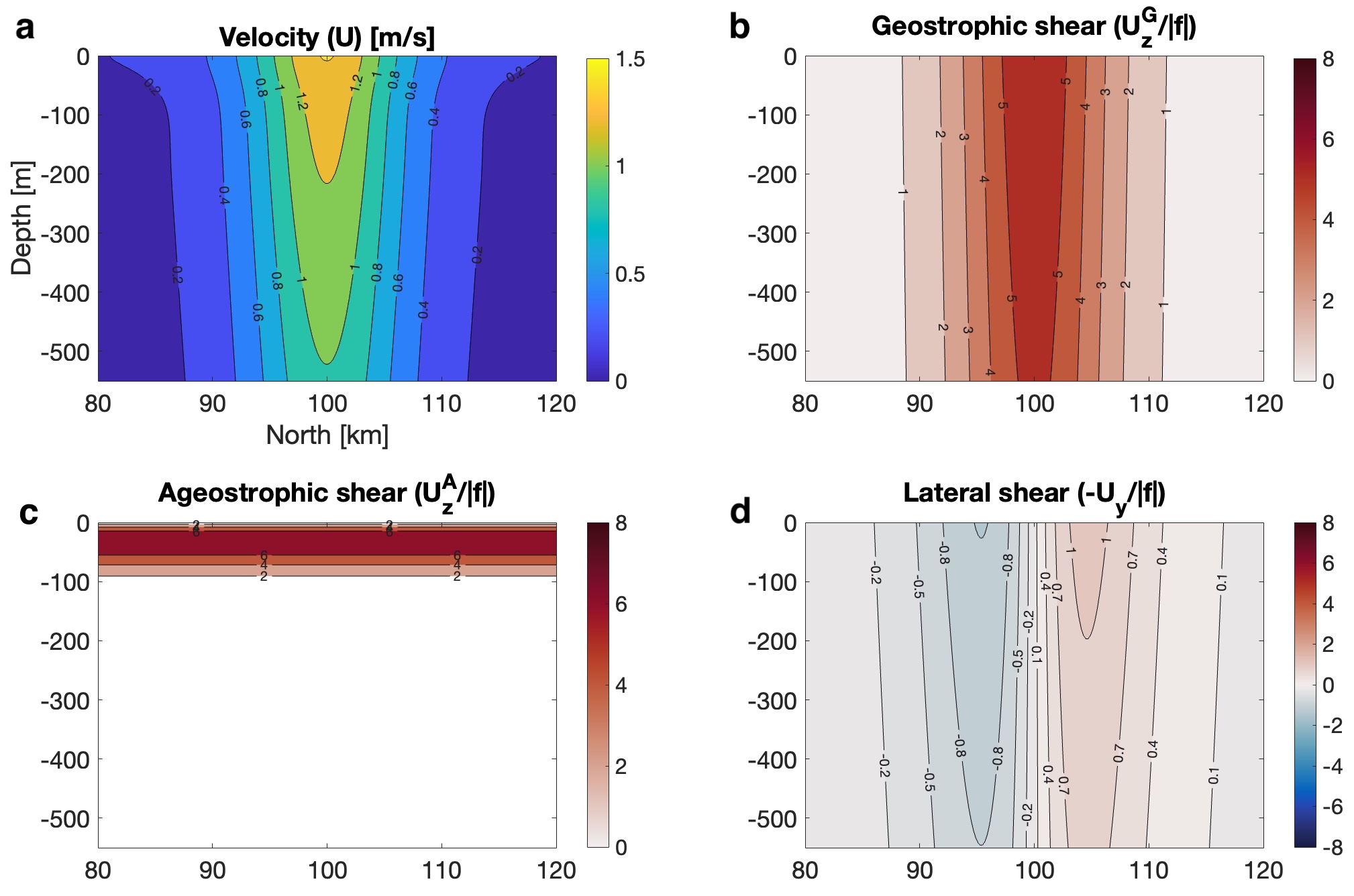}
    \caption{Showing 2-D model of an idealized geostrophic jet subjected to Ekman forcing. Here the ageostrophic velocity shear from an along-front Ekman transport (wind direction $\theta=270^\circ$ producing along-front net transport) is depicted in \textbf{(c)}.}
    \label{fig:3}
\end{figure}
 We apply the ageostrophic (AOI) criterion (Eq. \ref{eq:10}), using an idealized steady 2-D analytical feature model of an ACC zonal jet \cite<see>{Gangopadhyay02,Ferris25}, based on jet characteristics observed during a long-duration glider program \cite{Ferris22a,Ferris22b}. This flow is shown as an across-front slice of the jet in Fig.~\ref{fig:1}ab depicted in Fig. \ref{fig:2}. It is oriented in the zonal direction, consistent with the mean direction of the ACC. Distinct from the open ocean jet in \citeA{FerrisGong25}, we impose an idealized Ekman flow as an example of forced ageostrophy (Fig. \ref{fig:3}):
\begin{figure}[htbp]
    \centering
    \includegraphics[width=0.9\textwidth]{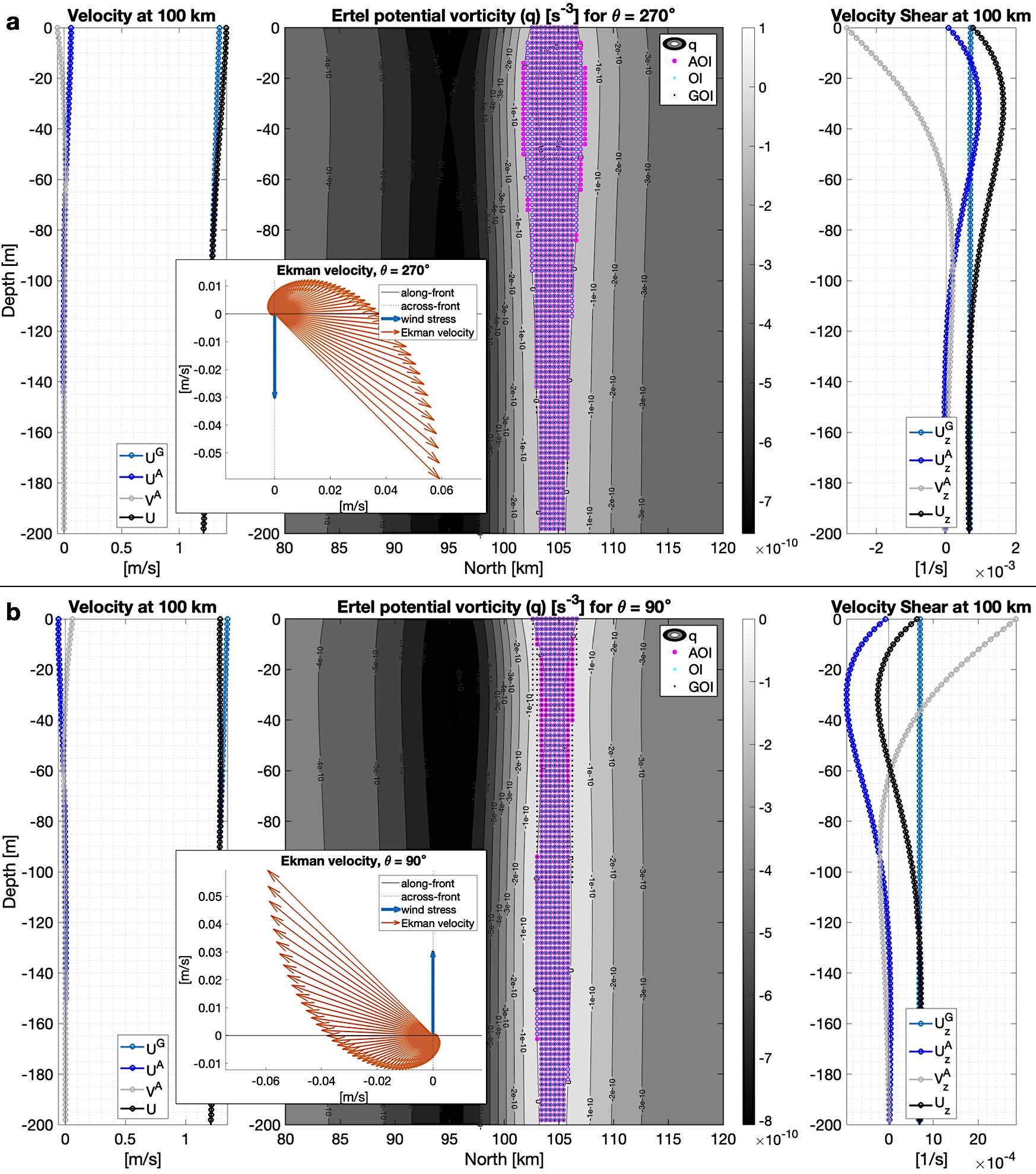}
    \caption{Showing velocity and vertical velocity shear profiles for the idealized jet in Fig. \ref{fig:2}, with overturning instabilities (Table \ref{tab:table3}) highlighted in the middle panel. The inset panel shows the ageostrophic velocity spiral produced by (Eq. \ref{eq:14}) relative to the zonal front.}
    \label{fig:4}
\end{figure}
\begin{figure}[htbp]
    \centering
    \includegraphics[width=0.9\textwidth]{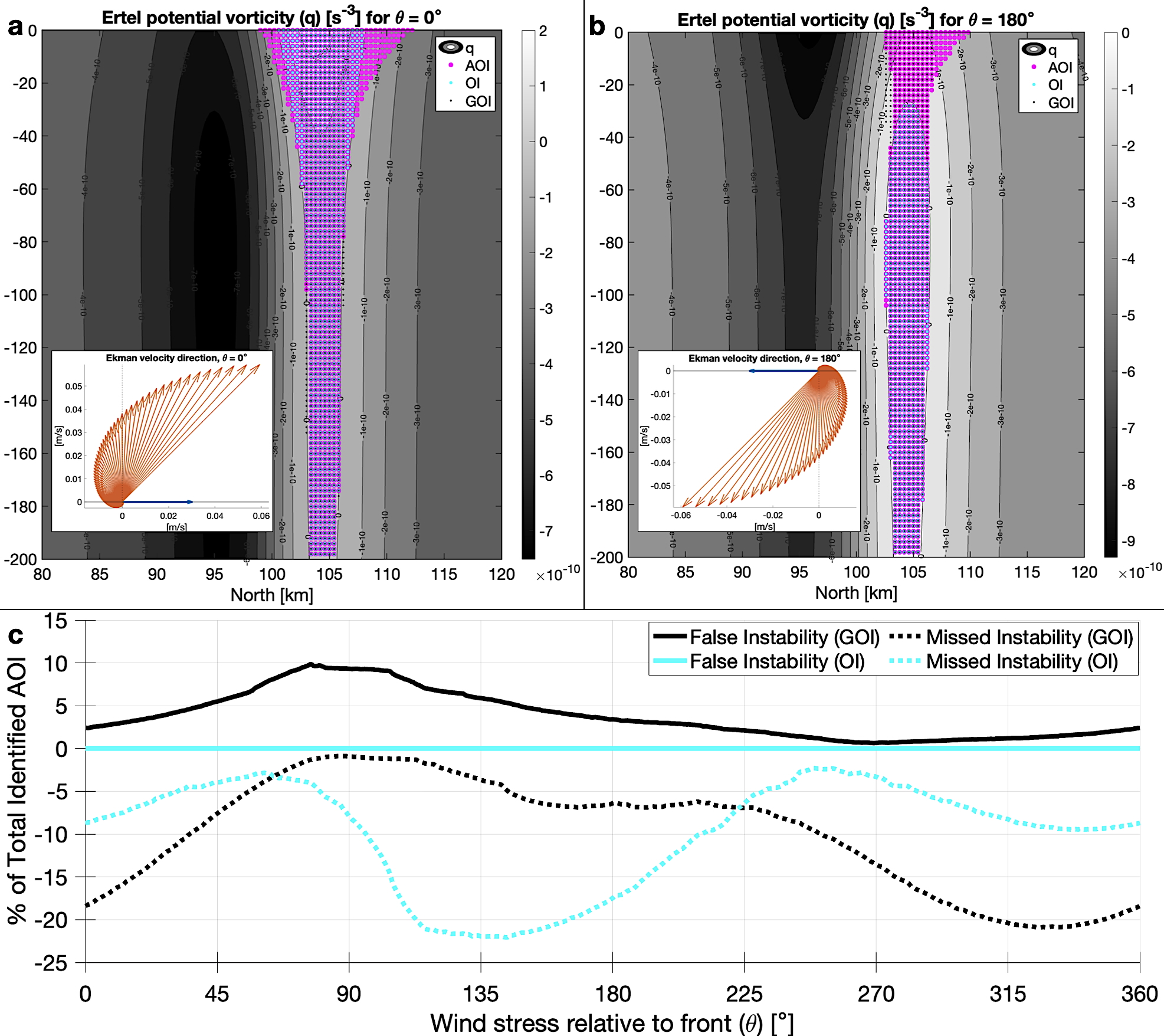}
    \caption{Showing \textbf{(a,b)} jet cross-sections as in Fig. \ref{fig:4} for additional wind angles and \textbf{(c)} the amount of false instability (solid lines) and missed instability (dotted lines) for the full range of wind angles when using GOI and OI criteria, which do not explicitly account for (de)stabilization by ageostrophic shear. False instability is defined as nodes where GOI/OI indicate instability but AOI does not, whereas missed instability is defined as nodes where AOI indicates instability but GOI/OI do not.}
    \label{fig:5}
\end{figure}

\begin{subequations}
\begin{equation}
U^A(z,y)=  \frac{\delta}{\rho_0\nu_E\sqrt{2}}e^{z/\delta}(\tau^x\cos\phi - \tau^y\sin\phi)
\end{equation}
\begin{equation}
V^A(z,y)=  \frac{\delta}{\rho_0\nu_E\sqrt{2}}e^{z/\delta}(\tau^x\sin\phi + \tau^y\cos\phi)
\end{equation}
\label{eq:14}
\end{subequations}
where $\phi = \text{sgn}(f) (z/\delta-\pi/4)$, and an Ekman depth of $\delta = \sqrt{2\nu_E/|{f}|} = 40$ m results from uniform surface wind stress $\tau= 0.3$ N m$^{-2}$ with an across-front wind direction $\theta$ such that $(\tau^x,\tau^y)= (\tau\cos{\theta},\tau\sin{\theta})$, and eddy viscosity $\nu_E = 10^{-1} $ m$^2$/s.

Overturning instabilities are identified (Fig. \ref{fig:4}) using GOI, OI, and AOI criteria (Table \ref{tab:table3}). The AOI criterion (Eq. \ref{eq:10}) essentially represents the augmentation of the total along-front velocity shear (black) by the ageostrophic along-front velocity shear (cobalt blue), resulting in identification of AOI (magenta nodes in Fig. \ref{fig:4}). There are no instances of pure SSI ($Ri < 0.25$) in these scenarios (Fig. \ref{fig:3}, Fig. \ref{fig:4}, Fig. \ref{fig:5}). The (de)stabilizing effects of velocity and/or velocity shear in the across-front direction (gray profiles in Fig. \ref{fig:4}) are neglected by all criteria. The top panel of Fig. \ref{fig:4} ($\theta = 270^\circ$, along-front positive Ekman transport) illustrates a case where forced ageostrophic shear combines with geostrophic shear to destabilize the flow such that AOI $>$ OI $>$ GOI. The bottom panel of Fig. \ref{fig:4} ($\theta = 90^\circ$, along-front negative Ekman transport) illustrates a case where ageostrophic shear predominately stabilizes the flow, such that there is false GOI (black dots alone). Criteria both for OI and AOI represent some of this stabilization by ageostrophic shear, with the OI criterion representing stabilization of the flow more aggressively (due to absence of the positive-definite parallel shear term in Eq. \ref{eq:10}, but not in $qf<0$, which drives $A$ towards instability).

We further illustrate the variation between GOI, OI, and AOI criteria by varying the flow via rotation of the Ekman velocity spiral counterclockwise 0-360º relative to the direction of the geostrophic frontal jet (Fig. \ref{fig:5}). The net effect of ageostrophic shear is not uniform with depth (see concurrent missed and false instability via the GOI criterion) and depends on the case-specific vertical profile of the along-front shear; illustrating how, in the presence of significant forcing, submesoscale instabilities such as SI may not act over a monotonic layer as they do in purely geostrophic flows (e.g. \citeA{Bachman17}). The parameterization of these instabilities in heavily forced regions (such as the subpolar oceans) may require a depth-varying methodology. It is our observation that the OI criterion misses a similar amount of instability as the GOI criterion, but is advantageous in that it does not produce false instability (in other words, is overly conservative); the latter of which was pointed out by \citeA{Haney15} in the context of wave effects.

 \subsection{Application to a RANS-type regional model}
 We now examine the AOI criterion using the Reynolds-averaged Navier-Stokes (RANS)-type model output examined by \citeA{Ferris24a} and archived online \cite{Ferris25}. The 1-km, 50-layer ROMS \cite{Shchepetkin05} model was initialized using the 1/12$^\circ$ resolution Operational Mercator and GLORYS12V1 operational simulation products. Flux forcing is calculated using bulk formulae \cite{Fairall96,Large81} using the evolving model state and specified atmospheric state obtained from MERRA \cite{Gelaro17}. The model uses the GLS vertical mixing parameterization \cite{Warner05} for vertical turbulent mixing of momentum and tracers.
\begin{figure}[htbp]
    \centering
    \begin{minipage}[b]{0.8\textwidth}
        \includegraphics[width=\linewidth]{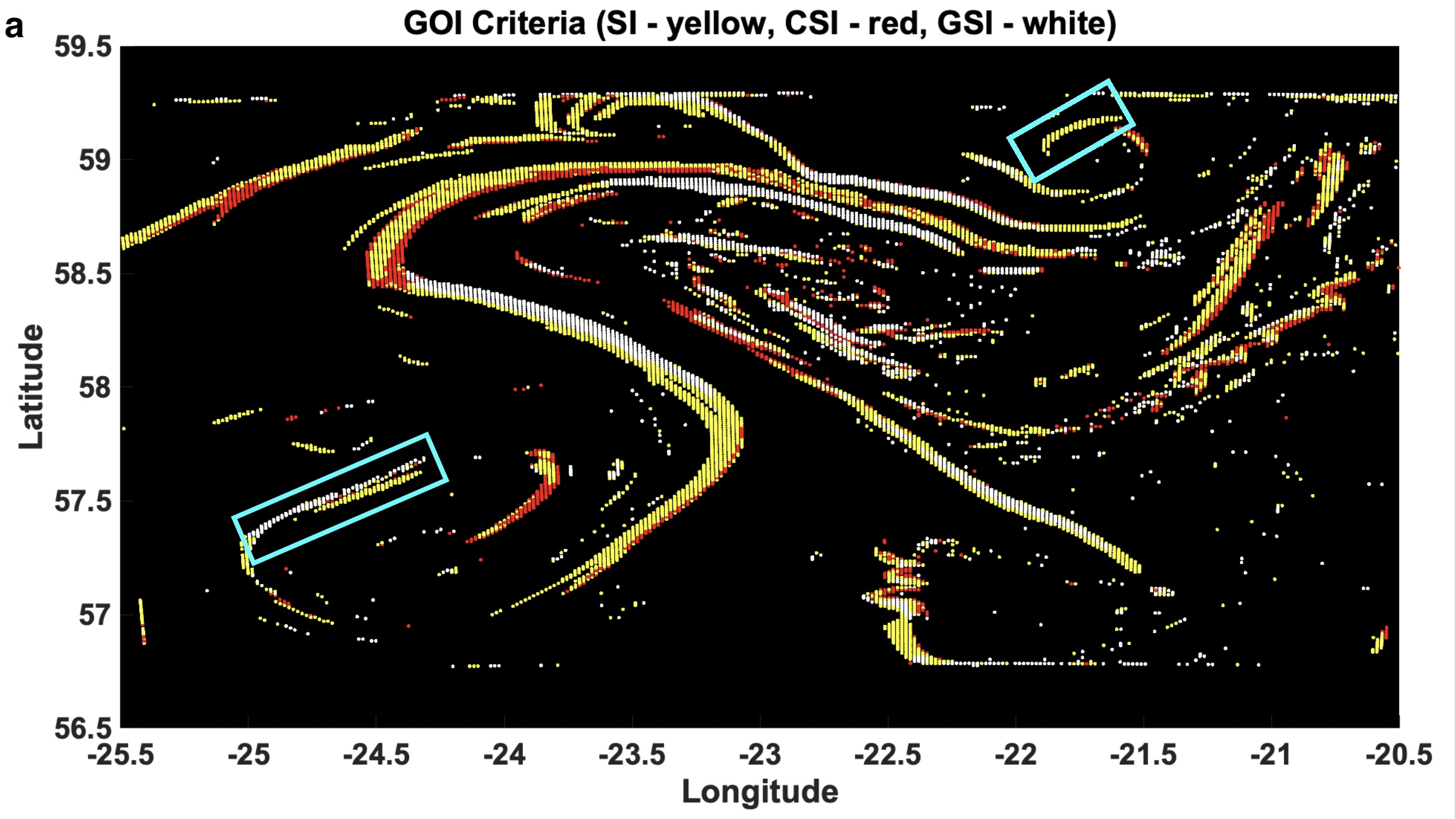}
        \label{fig:6a}
    \end{minipage}
    \begin{minipage}[b]{0.8\textwidth}
        \includegraphics[width=\linewidth]{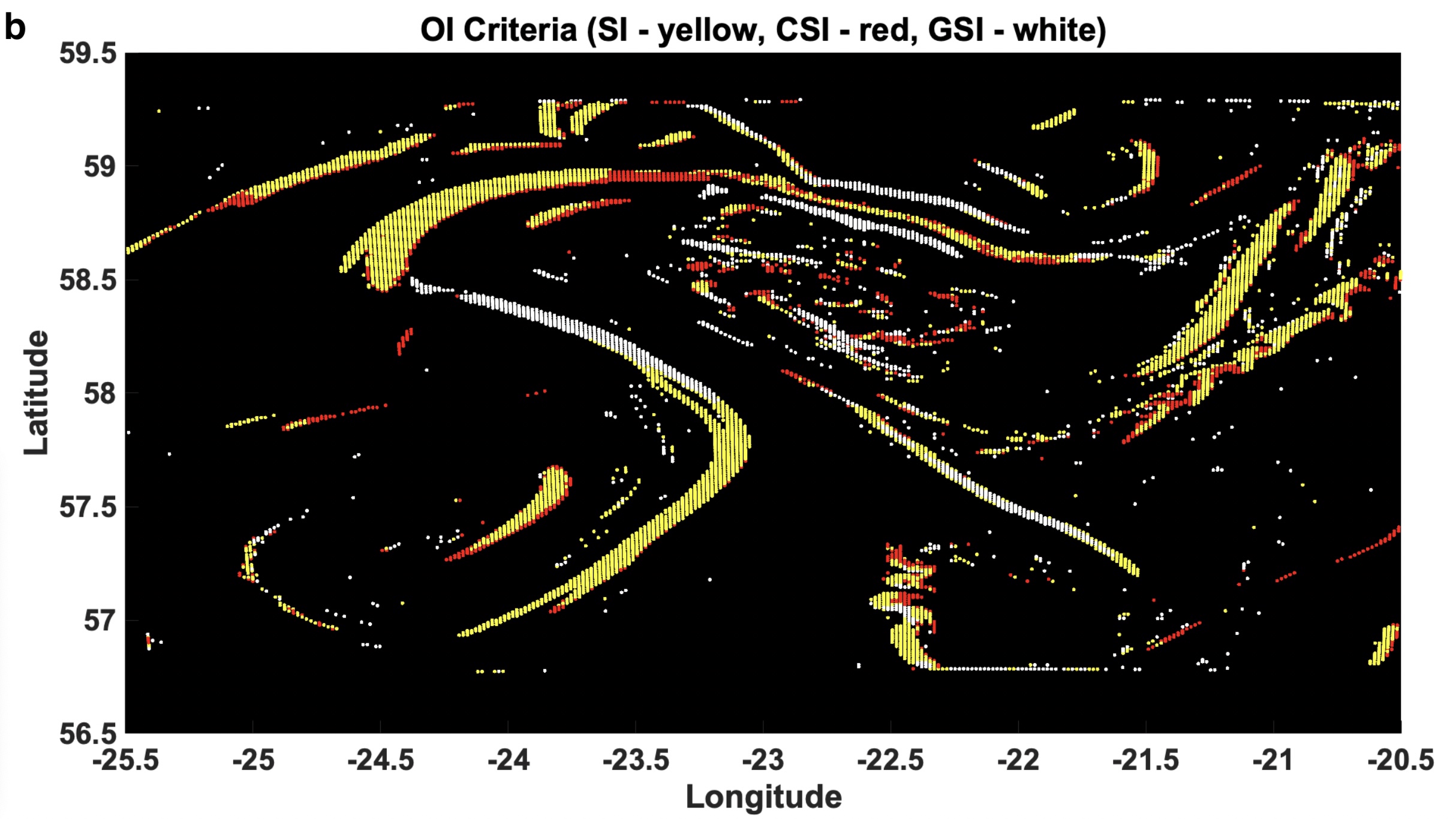}
        \label{fig:6b}
    \end{minipage}
    \begin{minipage}[b]{0.8\textwidth}
        \includegraphics[width=\linewidth]{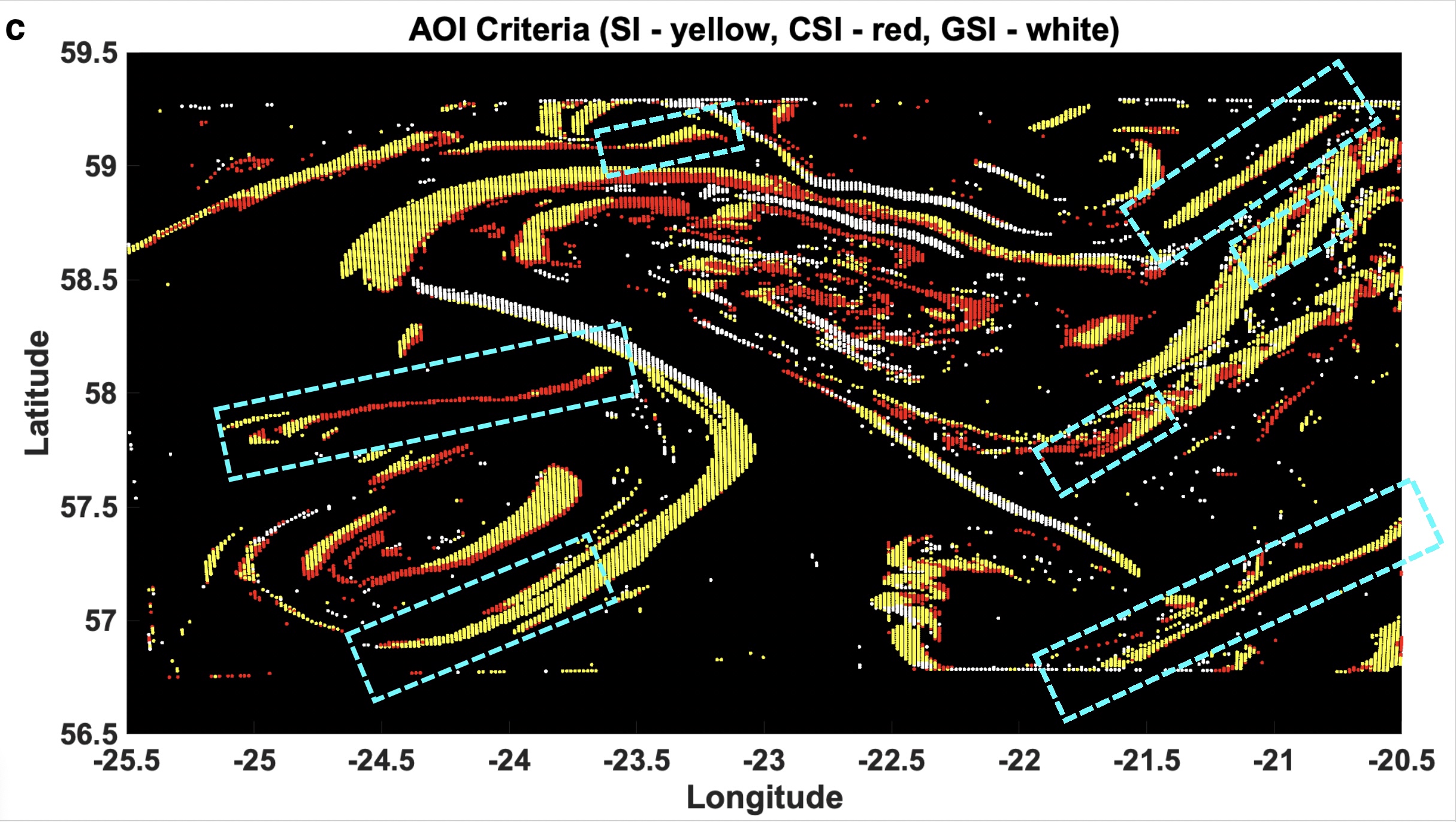}
        \label{fig:6c}
    \end{minipage}
    \caption{Top-down view of unstable structures identified in a 1-km ROMS simulation for one timestep (26-May-2018 08:00) of the NISKINe region $[^\circ]$ using \textbf{(a)} GOI, \textbf{(b)} OI, and \textbf{(c)} AOI criteria (Table \ref{tab:table3}); colored by instability subtype (Table \ref{tab:table4}). Nodes where $0 < Ri < 0.25$ have been omitted to avoid occlusion of frontal structures by mixed shear-symmetric instability at the air-sea interface. False instability is annotated with solid boxes and unstable fronts unique to AOI criteria (missed by GOI and OI criteria) are annotated with dashed boxes.}
    \label{fig:6}
\end{figure}

Resolving a submesoscale feature (e.g., a symmetrically unstable front) is different from resolving the instability or eddy that grows from it (were model resolution sufficient to let it progress through forward energy cascade). ROMS does not resolve the growing instability; rather it resolves the unstable setup (which is also the basis for subgrid scale parameterizations). To resolve these instabilities would require higher resolution; SI has wavelength 20-500 m \cite{Dong21a}. It is important to ensure we attain comparable resolution and gradients to previous observational and model-based studies of SI. Observations of symmetrically unstable features in the literature range 0.5-km to “few kilometer” towed CTD casts paired with shipboard ADCP \cite{DAsaro11,Thomas13,Thomas16,Adams17}. Lateral buoyancy gradients used in LES studies have ranged from $M^2 = B_x \approx 1.3 \times10^{-7}\text{s}^{-2}$ to $5 \times 10^{-7}\text{s}^{-2}$ \cite{Thomas13,Thomas16,Skyllingstad20} which are achieved in the simulations reported here. Another consideration is that ROMS is hydrostatic; however, hydrostacy is not expected to impact resolution of the unstable setup \cite{Mahadevan06}. A final consideration of linearization (Eq. \ref{eq:2}) is that we are assuming the unstable flow state (as identified using Table \ref{tab:table3} criteria does not represent instability in a state of mid-development. This is not an issue here given the model resolution is much larger than the wavelength of SI, but it is worth noting for future applications as ocean circulation model grid spacings decrease. 

We emphasized that the presence of AOI in the ROMS models (or any inviscid model) does not guarantee that submesoscale overturning instabilities will grow on a meaningful timescale; in the real ocean, viscous dissipation restricts the unstable wavenumber range by damping growth of the symmetric mode \cite{Ferris24a}. Additionally, across-front flow (e.g., Ekman buoyancy forcing) can advect light water over dense water, stabilizing the water column before meaningful growth of an instability --- a case which violates the frozen flow hypothesis. Another limitation of using a realistic simulation is that much like observations of the real ocean, to which the similarly curvature-neglecting $qf<0$ criterion is often applied, the model contains both time evolution and curvature of the base state. This is an unavoidable tradeoff of using a realistic ocean model to assess the real implications AOI over a large spatial scale and strongly motivates the expedient adaptation of (Eq. \ref{eq:10}) to include curvature and/or evolution.

\begin{figure}[htbp]
    \centering
    \includegraphics[width=0.7\textwidth]{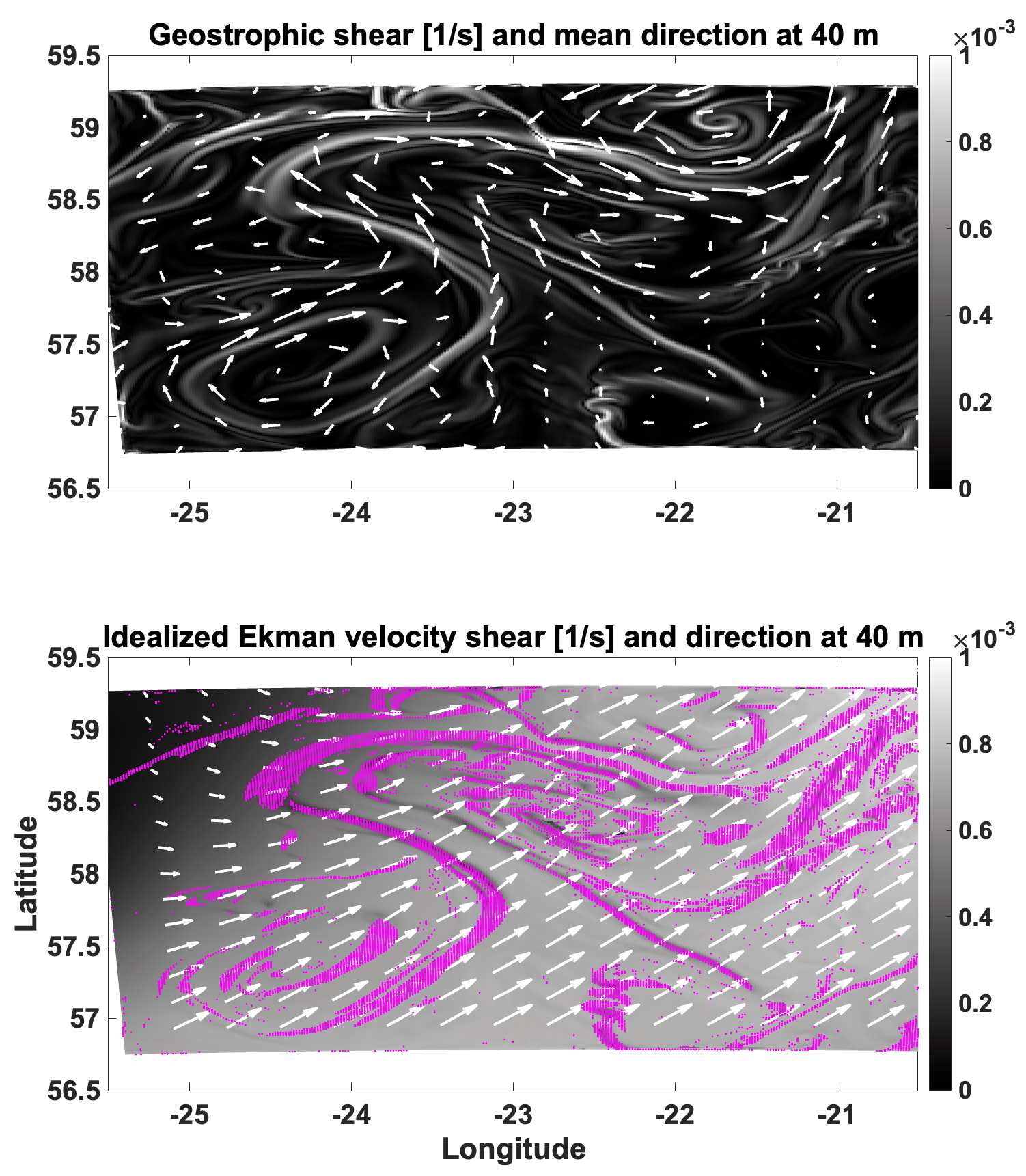}
    \caption{Showing for Fig. \ref{fig:6} a depth level of \textbf{(a)} geostrophic shear and \textbf{(b)} idealized Ekman velocity shear using Eq. \ref{eq:14}, with instabilities identified using AOI criteria. Wind stress is generally to the north, producing an expected Ekman flow to the north-east.}
    \label{fig:7}
\end{figure}

To demonstrate the differences between the GOI, OI, and AOI criteria (Table \ref{tab:table3}), we apply each criterion to a single timestep of model output for the NISKINe/high North Atlantic (Fig. \ref{fig:6}) region. Here we have omitted instances where $0<Ri<0.25$. At a basin scale, the criteria are similar. However, their representation of instability differs on a regional scale, where unstable ageostrophic filaments which are resolved by the model but missed by GOI criteria. The spatial distribution of instability is similar for the three criteria (Fig. \ref{fig:6}), but their differences can result in misidentifying the stability status of individual fronts; note that missed and falsely identified fronts have the same orientation; consistent with the influence of forcing direction (Fig. \ref{fig:7}). 

While the GOI criteria have practical applications in basin-scale analysis, the consideration of forced ageostrophic shear is critical to interpretation of regional or process-scale observations or model output. A task for future work is to understand the energetic significance of AOI in the upper ocean; the ROMS model is useful because it has broader and more realistic coverage than an idealized jet in an LES, but unlike an LES can only resolve the unstable setup rather than the growing instability itself. Without explicitly resolving the growing instability, we cannot admit AOI-scale turbulent fluxes $\overline{\bold{u}w}$ and leave this as an open question.
 
\section{Discussion}

In this study we have extended geostrophic methodologies \cite{Hoskins74,Thomas13} by presenting criteria (Eq. \ref{eq:10} and Eq. \ref{eq:13}b) that represent the stability-modifying effects of along-front ageostrophic shear in near-balanced flows. With LES and RANS-type models, three-dimensional fields of $U,V,B$ are available such that ageostrophic characteristics can and should be incorporated. We highlight the task of developing an AOI criterion which considers time evolution and/or curvature of the base state as an open challenge. A second open challenge is estimating the global energy contribution of AOI. A third open challenge is determining the relative role of SI and SI-type AOI in modifying upper ocean structure. Fueled by low stratification, high lateral shear, and/or high vertical shear, SI and its hybridized subtypes are often collocated with convective instability, centrifugal instability, and shear instability; the latter of which is not described by the CI-CSI-SI-GSI-GI framework (Table \ref{tab:table4}).

The significance of AOI is that its injects additional shear into the upper ocean, a process absent from RANS-type models below the resolved scales. The task of developing appropriate AOI parameterizations depends both on accurately identifying instability and understanding its energetics. Ageostrophic shear modifies the stability of the upper ocean, combining with background geostrophic shear to stabilize or destabilize the flow to SI and its hybridizations. Importantly, we have illustrated (Fig. \ref{fig:4} and Fig. \ref{fig:5}) that the setup of SI-type AOI is more complicated than previously theorized; it is not vertically uniform, and is dependent on the specific dynamic process that contributes ageostrophic shear. In heavily forced frontal zones, there is no "SI convective layer" in many circumstances. SI-type AOI may have significant levels of ageostrophic shear production, which must be considered as the community develops mixing parameterizations for SI and SI-type AOI in circulation models. This is not the first instance of advocacy for ageostrophic dynamics in parameterization: \citeA{Skyllingstad17} advocate for the importance of ageostrophic shear production in a front with destabilizing winds, and possible shortcomings of SI parameterizations based on geostrophoc shear production \cite{Bachman17,Dong21a}. \citeA{Haney15} similarly address SI influenced by Stokes shear, demonstrating the modification of stability by ageostrophic dynamics. 

Ageostrophic theory for identifying AOI is also needed to improve our understanding of its phenomenology. A specific application of this updated theory is the growing interest in instabilities along boundaries \cite{NaveiraGarabato19,Wenegrat20,Yankovsky21,Chor24}. Flows strongly influenced by topography (e.g., topographic shearing or drag) are not in geostrophic balance, and we must consider ageostrophic flow properties when investigating their effects and developing appropriate parameterizations. The community's interest in developing realistic submesoscale instability parameterizations is strong motivation for further observations in regions with an energetic submesoscale to better constrain the energetic contributions, depth ranges, and restratification timescales associated with these instabilities in the real ocean.

\acknowledgments
We thank John Klinck, Caitlin Whalen, Elizabeth Yankovsky, Christian Buckingham, Jeff Carpenter, and Jeffrey Early for helpful commentary on drafts of this paper; especially John Klinck, who shared the technique of using 2-D analytical models to investigate instability in ACC jets. Ferris’s effort was supported under the Applied Physics Laboratory - University of Washington Science \& Engineering Enrichment \& Development Fellowship and under ONR Northern Ocean Rapid Surface Evolution (NORSE) DRI award N00014211270. We thank Harper Simmons and Thilo Klenz for sharing invaluable technical expertise; the former of whom led production of the archived ROMS output \url{https://doi.org/10.25773/2EWW-DZ52} used in this paper.

\bibliography{ferrisgong26}

\end{document}